\documentclass[12pt]{iopart}
\pdfoutput=1
\usepackage[utf8]{inputenc}

\usepackage[english]{babel}

\expandafter\let\csname equation*\endcsname\relax

\expandafter\let\csname endequation*\endcsname\relax

\usepackage{amsmath}
\usepackage{amssymb}
\usepackage{mathrsfs}
\usepackage{graphicx}
\usepackage{graphics}
 \usepackage[parfill]{parskip}				
 \usepackage{pstricks}
\usepackage{subfig}
\usepackage{enumerate}
\usepackage{xspace}
\usepackage{cite}
 
\newcommand{\ket}[1]{\ensuremath{|#1\rangle}\xspace}
\newcommand{\bra}[1]{\ensuremath{\langle #1|}\xspace}

\newcommand{\psh}[2]{\ensuremath{\langle #1|#2\rangle}\xspace}

\begin{document}

\title[The single-particle density matrix of a quantum bright soliton]{The single-particle density matrix of a quantum bright soliton from the coordinate Bethe ansatz}

\author{Alex Ayet$^1$ $^2$ and Joachim Brand$^1$}
\address{$^1$ Dodd-Walls Centre for Photonics and Quantum Technologies, Centre for Theoretical Chemistry and Physics, and New Zealand Institute for Advanced Study, Massey University, Private Bag 102904, North Shore, Auckland 0745, New Zealand}
\address{$^2$ \'Ecole Normale Sup\'erieure, 45 rue d'Ulm, 75005 Paris, France}
\ead{\mailto{alex.ayet@ens.fr}, \mailto{j.brand@massey.ac.nz}}

\begin{abstract}
We present a novel approach for computing reduced density matrices for superpositions of eigenstates of 
a Bethe-ansatz solvable model by direct integration of the wave function in coordinate representation. A diagrammatic approach is developed to keep track of relevant terms and identify symmetries, which helps to reduce the number of terms that have to be evaluated numerically. As a first application we compute with  modest numerical resources the single-particle density matrix and its eigenvalues including the condensate fraction for a quantum bright soliton with up to $N=10$ bosons. The latter are constructed as superpositions of string-type Bethe-ansatz eigenstates of nonrelativistic bosons in one spatial dimension with attractive contact interaction. Upon delocalising the superposition in momentum space we find that the condensate fraction reaches maximum values larger than 97\% in the range of particles studied. The presented approach is suitable for studying time-dependent problems and generalises to higher-order correlation functions.
\end{abstract}

This is an author-created, un-copyedited version of an published 
in Journal of Statistical Mechanics: Theory and Experiment.  IOP Publishing Ltd is not responsible for any errors or omissions in this version of the manuscript or any version derived from it. The Version of Record is available online at https://doi.org/10.1088/1742-5468/aa58ac. 

\maketitle

\section{Introduction}

The study of one dimensional interacting bosons is an exciting field of ultra cold atomic physics where experimental results can match the theoretical predictions with an impressive precision \cite{Cazalilla2011,Jiang2015,Fabbri2015}. In the past few years, quasi-one-dimensional atomic gases have been realized experimentally \cite{Armijo2010,Paredes2004,Kinoshita2004} with the possibility to tune the interaction strength between particles, allowing comparison with theory.

The Lieb-Liniger model, which describes point-like bosons in one dimension interacting via contact interactions, is exactly solvable using the Bethe ansatz \cite{Lieb,Lieb2}.
A wider class of Bethe-ansatz solvable models like the Heisenberg spin chain and the Hubbard model at half filling share a closely related mathematical structure of the eigenfunctions \cite{Sutherland2004}. Even though the Bethe ansatz provides the eigenfunctions of a given model Hamiltonian in explicit form, it is still very difficult to evaluate many quantities of interest including general correlation functions. 
The introduction of the inverse quantum scattering method and its evolution into the algebraic Bethe ansatz \cite{Korepin}, has enabled significant progress. In particular the norm of the eigenstates \cite{Korepin1982} and certain one- and two-particle correlation functions of the ground state \cite{Caux} could be computed.

The coordinate Bethe ansatz refers to the original formulation of the Lieb-Liniger model, where the eigenstates were provided in coordinate representation \cite{Lieb,Lieb2}. Although any correlation function can be expressed by integrals over the eigenstates, the exponentially large number of terms often proves prohibitive beyond a very small number of particles. Only in very specific cases could closed form expressions for correlation functions be found. Examples for bound states of attractively interacting bosons are the particle density under the condition of a fixed center of mass position \cite{Calogero1975,Castin2008} and the particle density of a superposition of bound states \cite{Lai}. In this paper we compute the full single-particle density matrix of a superposition of bound states, which requires more general form factors than computed previously in \cite{Lai}, or by means of the algebraic Bethe ansatz in \cite{Caux}. We note that the coordinate Bethe ansatz has recently been used to calculate correlation functions for the repulsively interacting Bose gas in \cite{Zill2016}.

In comparison with the 1D Bose gas with repulsive interactions, much less attention has been paid to attractive interactions. Experimental realisations with ultra-cold atoms \cite{Nguyen2014,Medley2014, Marchant2013,Cornish2006, Khaykovich2002} have mostly been interpreted in terms of the Gross-Pitaevskii mean field theory, which leads to a cubic nonlinear Schr\"odinger equation, while some
theoretical works \cite{PhysRevLett.102.010403, PhysRevLett.94.090404, PhysRevLett.103.013902}
highlight the interesting physics that remains to be studied. The particularity of this regime is that the eigenstates of the system inlcude bound states \cite{McGuire1964} that behave like particles themselves. Although these states are delocalised in space, localised states can be constructed by proper superpositions. Such localised superpositions are quantum bright solitons, which exhibit a peak in their density, and have been studied in the past \cite{Lai}, even though only their density profile has been computed. They differ from the classical soliton solutions of the non-linear Schr\"odinger equation \cite{ZS} in their time evolution. Indeed, the center of mass of a quantum soliton will spread over time in order to recover the translational invariance of the system. Unlike the classical soliton, it does not conserve its shape over time. Other features are yet to be studied, such as their collision properties and  high-order correlation functions. This study is of  interest as bright solitons in Bose-Einstein condensates have been realized experimentally \cite{Nguyen2014,Khaykovich2002}, and understanding the features that the Gross-Pitaevskii approximation is missing is thus of special importance.

In this article we present a method to compute the full single-particle density matrix of such a solitonic state, using a similar starting point to the one in \cite{Lai}. While closed form expressions can be easily derived for the diagonal elements, the complexity of the integrals that need to be computed for the off diagonal elements requires a numerical implementation. We introduce a diagrammatic representation of these that simplifies the manipulation and allows for a very efficient computation of the relevant form factors.
On the one hand the approach yields new results such as the full density matrix, which gives access, via diagonalisation, to the condensed fraction of the quantum bright soliton. 
On the other hand it is a method that can be easily generalised to provide access to higher order correlation functions or for the study of more complex dynamics.

The paper is organized as follows. We recapitulate the coordinate Bethe Ansatz formalism and the form of the eigenstates of the system in the attractive case in \sref{part1}, followed by a discussion on the solitonic state we are studying in \sref{parts}.  After explaining the approach for the computation the density matrix in sections \ref{part2} and \ref{diff}, we discuss numerical results in \sref{finalpart}.

\section{The coordinate Bethe Ansatz formalism}\label{part1}

The Lieb-Liniger model represents a one dimensional system of $N$ bosons with coordinates $x_i$ interacting with a contact potential of strength $c$. The Hamiltonian of the model is
\begin{equation}\label{hamiltonian}
H =  - \sum_{i = 1}^{N} \frac{\partial^{2}}{\partial x_{i}^{2}} + 2c \sum_{\langle i , j \rangle} \delta (x_{i} - x_{j}), 
\end{equation}
where the sum runs over all  pairs of particles.

The coordinate Bethe Ansatz gives an explicit form for the eigenstates of the Schr\"odinger equation associated with the Hamiltonian \eref{hamiltonian}, in the position representation. If we denote  the size of the system by $2L$, we can write the eigenstates in the \emph{fundamental domain} $-L \leq x_1 \leq x_2 \leq ... \leq x_N \leq L$ in the form
\begin{equation}\label{ansatz}
 \psh{\{ x_{i} \}}{\{k_i\}} = \sum_{\mathcal{P} \in S(N)} a(\mathcal{P}) \exp\left(\sum_{i} k_{\mathcal{P}(i)} x_i\right),
\end{equation}
where $S(N)$ is the set of all permutations on ${1,...,N}$ and the $k_i$ are called \emph{quasi-momenta}. They can be either real or complex, depending on sign of $c$. The wave function on the whole domain can be reconstructed by bosonic symmetry.

The contact potential in \eref{hamiltonian} imposes boundary conditions for $x_i = x_j$, and leads to constraints over the coefficients $a(\mathcal{P})$ that have to satisfy the following set of equations:
\begin{equation}\label{coef}
a(\mathcal{P'}) = \frac{k_{\mathcal{P}(i+1)}- k_{\mathcal{P}(i)} + ic }{k_{\mathcal{P}(i+1)}- k_{\mathcal{P}(i)} - ic } a(\mathcal{P}),
\end{equation}
where $\mathcal{P'}$ is derived from $\mathcal{P}$ by exchanging the $i$-th and the $(i+1)$-th component of the permutation.

In the following, we are interested in the attractive case characterized by $c$ being negative, in which the eigenstates of the system include the so called string states. In the regime of a large $L$ and finite $N$, their quasi-momenta are complex, and form strings along the imaginary axis. The ground state and the first excited states, denoted by $\ket{p}$, are made of one string centered on the real axis on a value $p$, and the quasi-momenta of the particles are then
\begin{equation}\label{sstate-pseudo}
k_j = p + i \frac{c}{2}(N-2j+1) + \delta_{j} \quad j \in \{1,..N\},
\end{equation}	
with $\delta_{j} \sim e^{-(cst)L}$ called the string deviations. In the following, we will neglect the exponentially small string deviations, which becomes exact in the limit of large box size $2L$. The formation of strings is regarded a hypothesis and unproven in the general case of a finite box size but known to be exact in the infinite volume limit \cite{Bethe1931,takahashi1971one,essler1992fine}.

The string states are eigenstates of the momentum operator of the whole system with eigenvalues $Np$, the total momentum of the string. It follows from \eref{coef} that all the coefficients $a(\mathcal{P})$ vanish except for the identity permutation, when using the string quasi-momenta of \eqref{sstate-pseudo}.

The final wave function for a simple string state \eqref{sstate-pseudo} \emph{valid inside and outside of the fundamental domain} is thus
\begin{equation}\label{bound}
\psh{\{x_i\}}{p} = \mathcal{N}\exp\left( ip \sum_{j=1}^{N} x_j + \frac{c}{2} \sum_{1\leq i \leq j \leq N} |x_j - x_i| \right).
\end{equation}
From this representation it is easy to see that the wave function of a string state has a very simple structure. In particular, it is the product of a term $\chi(R)$ that depends only on the centre-of-mass coordinate $R= N^{-1}\sum_{j=1}^{N} x_j$ and another term that depends only on the distances between particle coordinates, i.e.\ on relative coordinates. The latter part represents an exponentially localised bound state and it is independent of the string momentum. The centre-of-mass wave function $\chi(R) = \exp(i Np R)$ simply represents a plane-wave with momentum $Np$.
The string state \eqref{bound} is an eigenstate of the Hamiltonian \eref{hamiltonian} with energy
\begin{equation}\label{energy}
E_p = Np^2 - \frac{c^2}{12}N(N^2-1).
\end{equation}

We end this section by giving the normalization factor of the eigenstates for our spatially discretised numerical setup
\begin{equation}\label{Norm}
\mathcal{N} = \left(\frac{2NL}{c^{N-1}(N-1)!}\right)^{-1/2}.
\end{equation}
It is computed in detail in \ref{appen-norm}.

\section{Relation to the classical bright soliton}\label{classical}

A variational approximation for the ground state of the Hamiltonian \eqref{hamiltonian} is provided by the Hartree-Fock method \cite{Castin2001}. Starting from an ansatz for the many-body wave function as a product of $N$ identical single particle functions $\phi(x)$, the wave function $\phi(x)$ is obtained from minimising the Hartree-Fock energy functional
\begin{align}
E^\mathrm{HF}[\phi,\phi^*] = N \int d z \left[ \left|\frac{d \phi}{d z}\right|^2 + (N-1) c |\phi(z)|^4 \right] .
\end{align}
Requiring stationarity of  this functional yields the time-independent version of the Gross-Pitaevskii or nonlinear Schr\"odinger equation
\begin{align}
\mu \phi(z) = - \frac{d^2\phi}{dz^2} + \frac{2(N-1)c}{N}|\phi(z)|^2 \phi(z),
\end{align}
where the chemical potential $\mu$ is a Lagrange multiplier to assure the normalisation  $N= \int dz |\phi(z)|^2$  of the  Gross-Pitaevskii wave function $\phi(z)$. Taking the box size $L\to\infty$, the well-known solution for attractive interactions $c<0$ is the bright soliton 
\begin{align}
\phi^\mathrm{sol}(z) = \frac{\sqrt{N}}{2\sqrt{\xi}}\,\mathrm{sech}\left(\frac{z-z_0}{2\xi}\right).
\end{align}
A peculiar feature is that this solution and the associated particle number density $n(x) = |\phi^\mathrm{sol}(x)|^2$ are spatially localised  at an arbitrary position $z_0$, which represents an infinite degeneracy. This is in stark contrast to the exact quantum ground state, the string state with zero total momentum, which is completely delocalised. The localisation length scale $\xi= [(N-1)c]^{-1}$ corresponds to a variance of the number density \cite{Cosme2015a}
\begin{align}\label{eq:HFvariance}
\sigma_\mathrm{HF}^2 = \frac{\pi^2 \xi^2}{3} = \frac{\pi^2}{3 c^2 (N-1)^2} .
\end{align}

\section{The quantum bright soliton state}\label{parts}

The aim of this article is to compute the single-particle density matrix of a quantum bright soliton with a definite number of particles, constructed with Bethe eigenstates. Since the latter are completely delocalise, we will need to consider superpositions of Bethe eigenstates.

We want to construct a localized wave packet with massive particles, where quantum superpositions of different particle number, as previously considered in Ref.\ \cite{Lai}, are physically not meaningful \cite{Leggett2006}. It is thus natural to take a superposition of single string states of $N$ particles  with a Gaussian momentum distribution centered on $P_0 = \frac{\pi}{L}n_0 \in \mathbb{R}$ with width $\Delta$.  We are considering a box of size $2L$ with periodic boundary conditions, leading to a discretisation of the total momentum of the strings we are taking in the superposition. The superposed state we are considering is 
\begin{equation}\label{s-state}
\ket{S} =\mathcal{G}\sum_{n = n_0 -\frac{s}{2}}^{n_0 + \frac{s}{2}} \exp\left[-\frac{\pi^2}{L^2 \Delta}(n-n_0)^2 \right] \ket{\frac{\pi}{L}n} ,
\end{equation}
with $s$ being a threshold we include for our numerical implementation. Ideally, we would like to take it to infinity. $\mathcal{G}$ is a normalization factor for the discrete Gaussian distribution we are using.

Since the string states appearing in the sum in \eqref{s-state} only differ in the centre-of-mass part of the wave function [as seen from Eq.~\eqref{bound}] but share the same relative coordinate part, the superposition $\ket{S}$ still factorises according to centre-of-mass and relative coordinate dependence. The centre-of-mass wave function corresponds to a Gaussian wave packet, which is localised in space. The particle-number density, which is given by the diagonal part of the single-particle density matrix, $\rho(x,x)$, is fairly easy to compute \cite{Lai}. It can further be shown that the variance of the quantum soliton state $\ket{S}$ is the sum of variance of the centre-of-mass wave packet and the variance of the relative-motion wave function \cite{Cosme2015a}. Computing the off-diagonal parts of the single-particle density matrix is non-trivial and is the main subject of this paper. 

Let us now recall the expression of the single-particle density matrix (also called one-body density matrix) in its first quantized form. Given a state $\ket{S}$ of $N$ particles, it takes the form
\begin{equation}\label{densityf}
\rho(x',x) = N \int_{[-L,L]^{N-1}} dx_1...dx_{N-1 } \psh{S}{x_1,...,x_{N-1},x'}\psh{x_1,...,x_{N-1},x}{S}.
\end{equation}

In the case of the state \eref{s-state}, if we define a form factor as 
\begin{equation}\label{form}
\mathcal{F}_{p',p}(x',x) = \int_{[-L,L]^{N-1}} dx_1 ... dx_{N-1}  \psh{p'}{x_1,...,x_{N-1},x'}\psh{x_1,...,x_{N-1},x}{p},
\end{equation}
we obtain for the density matrix,
\begin{equation}\label{d-matrix}
\rho(x',x) = N\mathcal{G}^2 \sum_{n,n' = n_0 -\frac{s}{2}}^{n_0 + \frac{s}{2}} \exp\left\{-\frac{\pi^2}{L^2 \Delta}\left[ (n'-n_0)^2  + (n-n_0)^2 \right]\right\}\mathcal{F}_{\frac{\pi}{L}n' , \frac{\pi}{L}n}(x',x).
\end{equation}

The computation of the form factors requires a numerical implementation. Due to its complexity, it is the aim of the next section to give some details on it.

\section{Computation of the form factor}\label{part2}

In this section we give some details on the computation of the form factors for the single-particle density matrix of the solitonic state $\ket{S}$. 

The expression for the form factor \eqref{form} can be simplified by extending the integration domain to $\mathbb{R}^{N-1}$, which is justified by the exponential localisation of the expression \eref{s-state}. This assumption leads to an approximation at the same level as neglecting the string deviations in \eqref{sstate-pseudo}. It is valid in the regime  $\frac{1}{c}\ll L$.

Now, given the symmetry of the integrand, we can rewrite the integral \eqref{form} in the fundamental domain to obtain
\begin{multline}
\mathcal{F}_{p',p}(x',x) = (N-1)! \mathcal{N}^2 \sum_{0 \le m \le m' \le N-1} \int_{-\infty \le x_1 \le...\le x_m \le x \le x_{m+1} \le... \le x_m' \le x' \le x_{m'+1} \le ... \le x_{N-1} \le \infty} \\ dx_1 ... dx_{N-1}   \exp \left[i(px-p'x') +\frac{c}{2}(2m-N+1)x +\frac{c}{2}(2m'-N+1)x' \right]  \\\times \exp \left[ i(p-p') \sum_{j=1}^{N-1} x_j - c \sum_{j=1}^{N-1}(N-2j)x_j \right] \exp \left[ - c \sum_{j=1}^{m} x_j + c \sum_{j=m'+1}^{N-1} x_j \right],
\end{multline}
with the convention that $x_0 = - \infty$ and $x_N = \infty$.

Trying now to compute one of the integrals of the above sum, we notice that it is the product of three independent factors which are 
\begin{multline}\label{simple1}
 \mathcal{I}^{1}_{m}(x',x) = \int_{x_m = - \infty}^{x}...\int_{x_2 = -\infty}^{x_3}\int_{x_1 = - \infty}^{x_2} dx_1 ... dx_{N-1} \\ \times \exp\left[i(p-p') \sum_{j=1}^{m} x_j - c \sum_{j=1}^{m}(N-2j+1)x_j \right]
\end{multline}
\begin{multline}\label{simple2}
 \mathcal{I}^{2}_{m'}(x',x) = \int_{x_{m'+1} = x'}^{\infty}...\int_{x_{N-2} = x_{N-3}}^{\infty}\int_{x_{N-1}=x_{N-2}}^{\infty} dx_1 ... dx_{N-1} \\ \times \exp\left[ i(p-p') \sum_{j=m'+1}^{N-1} x_j - c \sum_{j=m'+1}^{N-1}(N-2j-1)x_j \right]
\end{multline}
\begin{multline}\label{complicated}
  \mathcal{I}_{m',m}(x',x) =\int_{x_{m'} = x}^{x'}...\int_{x_{m+2} = x}^{x_{m+3}}\int_{x_{m+1}= x}^{x_{m+2}} dx_1 ... dx_{N-1}
\\ \times \exp\left[ i(p-p') \sum_{j=m+1}^{m'} x_j - c \sum_{j=m+1}^{m'}(N-2j)x_j\right].
\end{multline}
It is worth mentioning here that the convergence of these integrals is ensured by the negativity of $c$.

The first two factors can be integrated in closed form, and give respectively
\begin{equation}
 \mathcal{I}^{1}_{m}(x',x) = \frac{1}{m!} \prod_{r=1}^{m} \frac{\exp\left[ im(p-p')x - cm(N-m)x \right]}{ - c(N-r) - i(p-p')} ,
\end{equation}
\begin{equation}
 \mathcal{I}^{2}_{m'}(x',x) = \prod_{r = 1}^{N-1-m'} \frac{\exp\left[ i(N-1-m')(p-p')x' + c(N-1-m')(m'+1)x' \right]}{[(N-1-m')!][c(N-r) - i(p-p')]} .
\end{equation}

The third factor is more complicated. When we compute one of the successive integrals in $\mathcal{I}$, we obtain two terms, corresponding to the upper and lower limit of the integration domain. In the case of the first two factors $\mathcal{I}^{1}$ and $\mathcal{I}^2$, one of these terms vanishes because of the limit being infinity, but for $\mathcal{I}$, none of these terms cancels, and they accumulate in the next integrals.

As a consequence, we recover $2^{m'-m}$  terms to compute. A numerical implementation is necessary in order to evaluate them. These computations are based on a diagrammatic representation of the integrals, which leads to a reduction of the computational time. The diagrams help sorting out the different terms, in order to factor out their position dependence in certain cases, decreasing the computational time. They are also the basis of the numerical implementation itself.

The diagrammatic representation, the key to the solution of the problem, are presented in the next section for the case of a form factor between two different eigenstates. The case of identical eigenstates is discussed in \ref{same}, as it requires extra care and is more complex.

\section{Form factor between two different eigenstates}\label{diff}

In this section, we give some details on the computation of the integral \eqref{complicated}.
Let us first set the following notation that will be also used in \ref{same}. We define the partial sum $u_{i,j} = c\sum_{k=i}^{j} (N-2k) $, which yields
\begin{equation}\label{udef}
u_{i,j} =\begin{cases}
  c (N-i-j)(j-i+1), &\mbox{if } i \leq j \\
			0, &\mbox{otherwise}.
	\end{cases}
\end{equation}
We also denote by $P = (p-p')$ the difference of the momenta between the two strings involved in the form factor we are considering.

The derivation of the form factors involves the computation of a sequence of integrals, each on a different variable, starting by integrating $x_{m+1}$ until $x_{m'}$. Each of these integrations gives rise to two terms,  one coming from the lower limit of the integral $x_i = x$, and the other one from $x_i = x_{i+1}$. Unlike the case of equations \eqref{simple1} and \eqref{simple2}, none of this terms vanishes. 

\begin{figure}[!h]
\centerline{
\subfloat[first term]{
	\includegraphics[scale = 0.2]{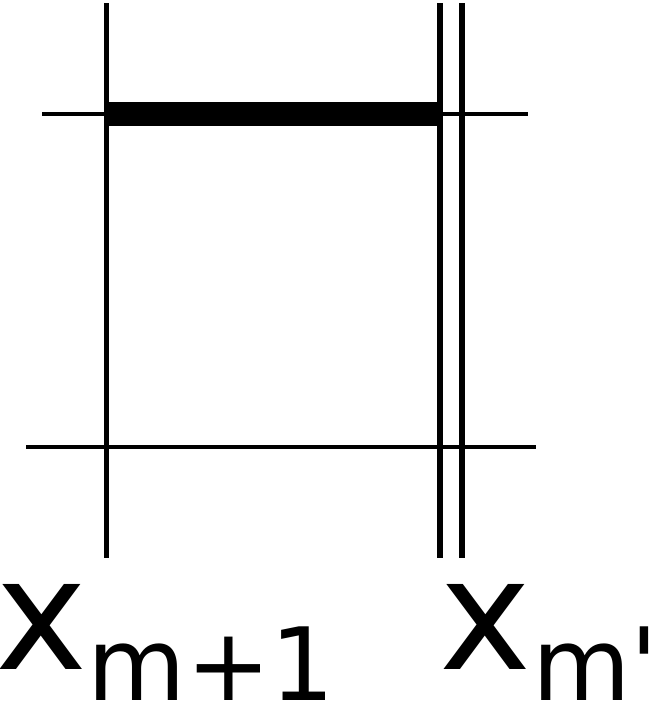}
	\label{1a}
} \;\;\;\;\;\;\;\;\;
\subfloat[second term]{
	\includegraphics[scale = 0.2]{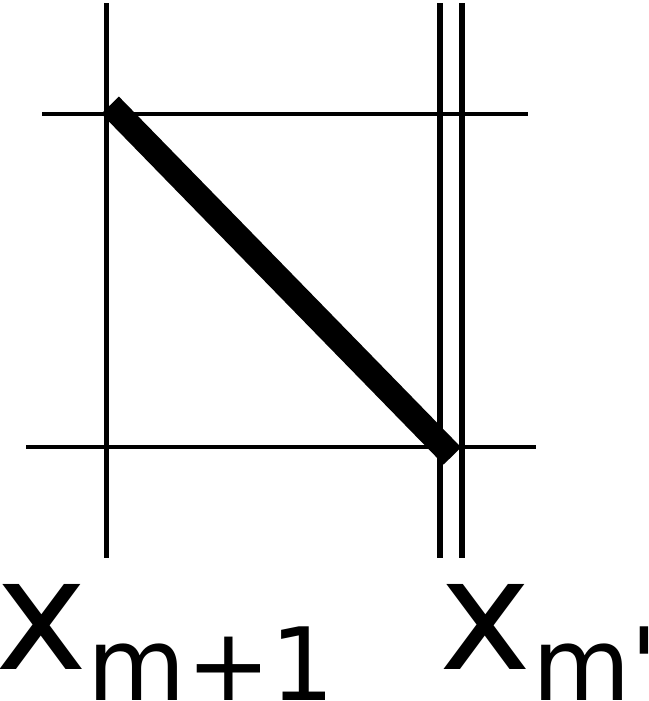}
	\label{1b}
}\;\;\;\;\;\;\;\;\;
\subfloat[third term]{
	\includegraphics[scale = 0.2]{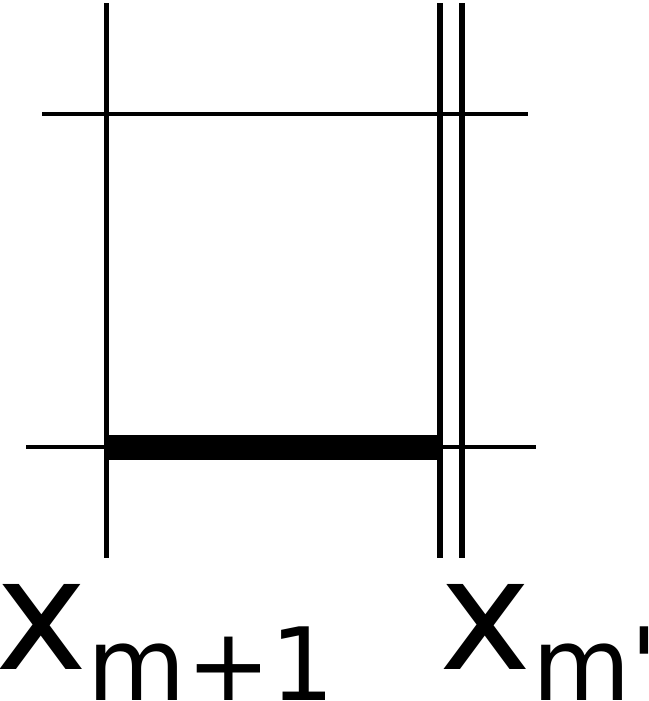}
	\label{1c}
}\;\;\;\;\;\;\;\;\;
\subfloat[fourth term]{
	\includegraphics[scale = 0.2]{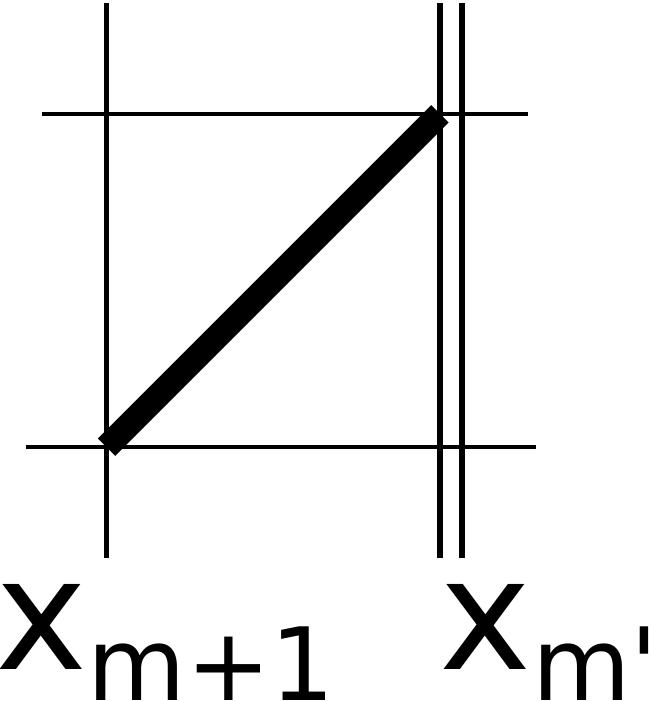}
	\label{1d}
}
}
\caption{Elementary diagrams, from which more complicated diagrams are constructed. The four diagrams also correspond to the four terms in \eqref{example} from left to right, respectively. Diagram \ref{1a} is a $2$-diagram, \ref{1b} and \ref{1c} are $0$-diagrams, and \ref{1d} is a $1$-diagram.}
\label{exdiag}
\end{figure}

First, let us consider the simple case of $m' = m+2$ in which we must integrate only over two variables. When we perform the first integral, we obtain
\begin{multline}\label{ex1}
 \mathcal{I}_{m',m'-2}(x',x) = \int_{x}^{x'}  dx_{m'}\frac{\exp\left[(iP - u_{m',m'})x_{m'}\right]}{iP - u_{m'-1,m'-1}}\\ \left\{ \exp\left[(iP - u_{m'-1,m'-1})x_{m'}\right] -  \exp\left[(iP- u_{m'-1,m'-1})x\right] \right\}.
\end{multline}
We have now two integrals to compute, giving us four terms. To obtain one of these terms, we must have first chosen one of the two integrals in \eqref{ex1}, and then either the term associated to the $x'$ limit or to the $x$ limit in the remaining integral. This succession of choices can be represented as a diagram. For the simple case of \eqref{ex1}, these are presented in \fref{exdiag}, and are associated to the four terms in the following equation
\begin{multline}\label{example}
\mathcal{I}_{m',m'-2}(x',x) = \frac{1}{iP - u_{m'-1,m'-1}}\left\{ \frac{\exp\left[(2iP - u_{m'-1, m'})x'\right]}{2iP - u_{m'-1,m'}} - \frac{\exp\left[(2iP - u_{m'-1,m'})x\right]}{2iP - u_{m'-1,m'}} \right. \\ -  \frac{\exp\left[(iP - u_{m',m'})x'+ (iP - u_{m'-1,m'-1})x\right]}{iP - u_{m',m'}} \left. + \frac{\exp\left[(2iP - u_{m'-1,m'})x\right]}{iP - u_{m',m'}}\right].
\end{multline}

For a more general $\mathcal{I}$, this representation consists in a $2 \times (m'-m)$ grid, where each column is one of the integrals, and the upper and lower rows represent the upper and lower limits of each integral. For instance, taking the upper point in column $i$ for a diagram means that for the integral on $x_i$, we choose the term corresponding to the upper limit $x_i = x_{i+1}$.
Note that the diagrams, unlike the integrals, are read from left to right, the first column on the left corresponding to the first integration. 

\begin{figure}[!h]
\centerline{
\subfloat{
 	\includegraphics[scale=0.4]{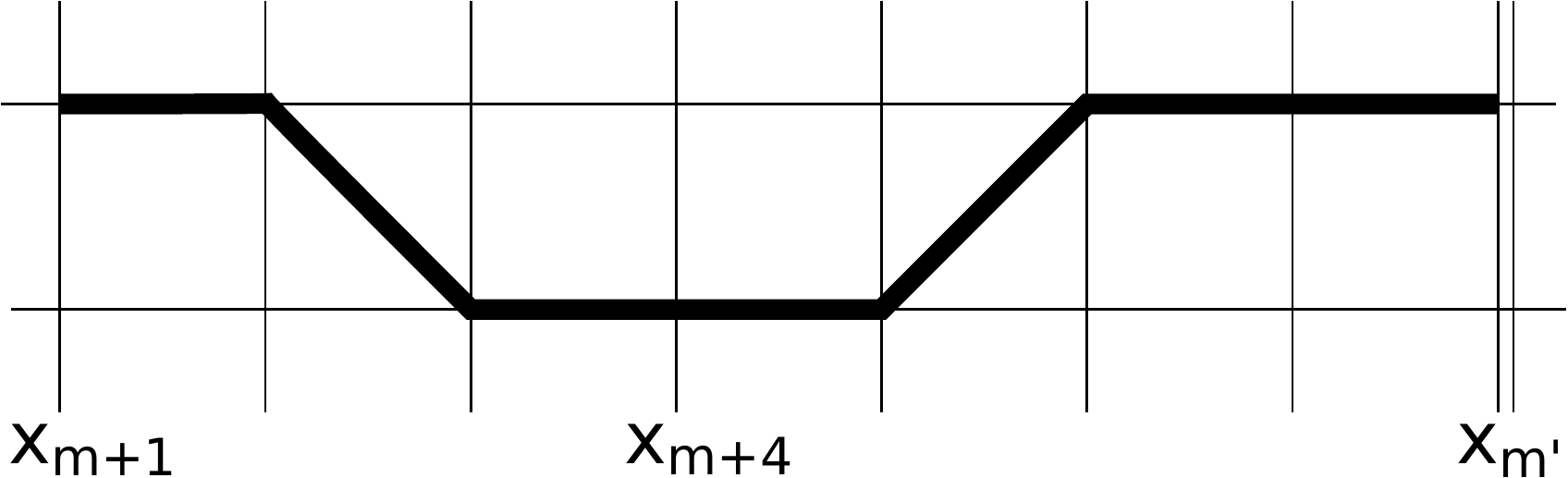}
}}

\centerline{
\subfloat{ 
	 \includegraphics[scale = 0.4]{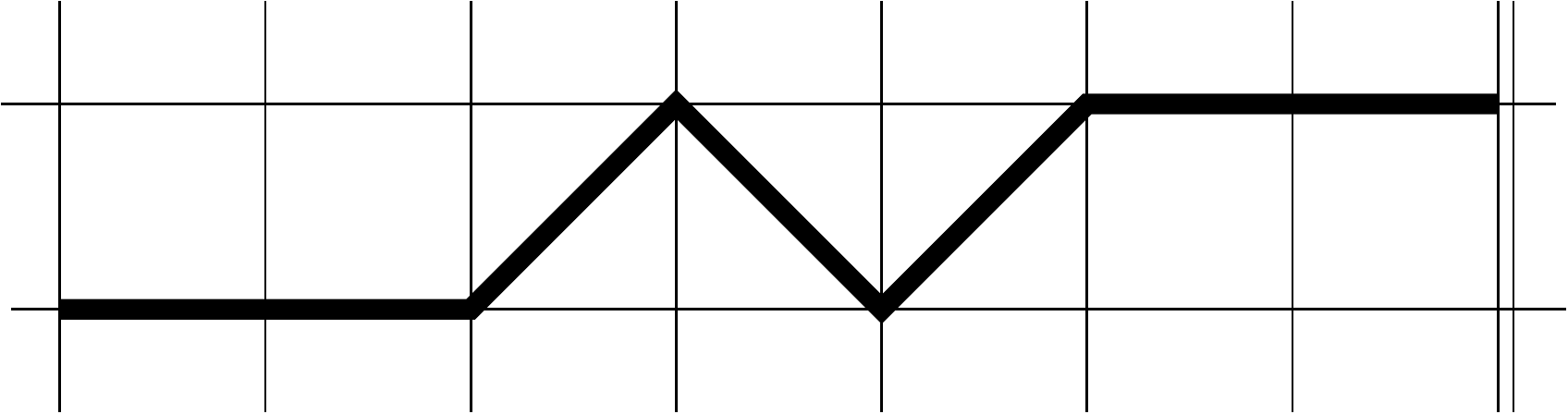}
}}
\caption{Two examples of 3-diagrams. They both carry the same exponential factor but not the same prefactor. The "fixed" part of the diagrams starts at column 6, while what happens before does not affect the exponential dependence of the whole diagram.}
\label{ldiag}
\end{figure}

This diagrammatic representation is useful in two ways. From one side, it is a basis for the code that computes this terms, which works constructing recursively all the possible diagrams. From the other side, it also diminishes computational time using the concept of $l$-diagrams. We can notice that, given $m$ and $m'$, each term of  $\mathcal{I}_{m',m}$ is of the form
\begin{equation}\label{l-diag}
\mathcal{C}e^{iP (lx' + (m'-m-l)x) - u_{m'-l+1,m'}x' - u_{m+1,m'-l}x},
\end{equation}
with  $l \in \{0,..,(m'-m) \}$. An $l$-diagram is defined as a diagram representing this type of contribution for a given $l$. Graphically, it is a diagram that has a number $l$ of ``upper" choices in a row before the last column (last column included), as shown in \fref{exdiag} and \fref{ldiag}.

To understand the latter form \eqref{l-diag} for the exponential dependence and its link to the diagrammatic representation, we must see that when computing an integral associated to a diagram, the result of the $j$-th integral (where we integrate over $x_{m +j}$) will depend on the choices we have done for the previous integrals. If, in one of the previous integrals, let's say when we integrate over $x_i$, we choose the upper limit term, then its dependence on $x_{i+1}$  will be carried over to the next integral, while the other choice would lead to a simple prefactor depending on $x$. We can thus notice a very important fact: \emph{whatever the choices for the previous integrals are, as soon as we choose a \emph{low} term (in the sense of the diagrams), all the choices we have made before this integration will not affect the $(i+1)$-th integral and the following integrals.} This is true in particular for the exponential dependence, which is the important part.

Now, the $x'$ dependence in \eqref{l-diag} can only come from the last integration, over the variable $x_{m'}$. It is thus only determined by the number of ``upper choices" in a row starting from the last column of a diagram. The exponential dependence in $x$ is determined before this succession of upper choices, and thus the final exponential dependence is determined by the number $l$ of upper choices we make in a row. Two examples of $3$-diagrams are presented in \fref{ldiag}.

We can now, instead of computing $\mathcal{I}$ by summing over all the diagrams, simply sum over the $l$-diagrams, i.e. factorise the terms that have the same exponential contribution. For a given $l$, we have to compute all the different possible paths until the fixed part starts (see \fref{ldiag}), because they all carry a different prefactor $\mathcal{C}$. This reduces the number of diagrams we have to compute. The huge advantage is that we are interested in computing the density matrix over all the space, and the position dependence only appears in the exponentials. This means that we can store the prefactors, computing them only once for the whole density matrix.

We give here a recursive procedure to compute the prefactor associated to a given diagram. We read the diagrams from left to right, and each step consists in moving froward of one column. Let $\mathcal{C}^{n}_{k}(P,N)$ be the prefactor at step $n$ for a given difference in momenta $P$ and a number of particles $N$ ($k$ being a memory variable of the number of ``upper" terms in a row before the actual step). Then $\mathcal{C}^{n+1}_{k'}(P,N)$ is obtained by a product
\begin{equation}
 \mathcal{C}^{n+1}_{k'}(P,N) = \mathbf{T}_{n+1 ,k}^{i} \mathcal{C}^{n}_{k}(P,N) \text{.}
\end{equation}
The factor $\mathbf{T}$ and the new memory variable $k'$ are obtained by identifying which of the four elementary diagram of \fref{exdiag} lies between the $(n+1)$-th and the $n$-th column:
\begin{equation} \label{diffrecursion}
\begin{array}{lll}
\mbox{(\ref{1a}), (\ref{1d}): }  \mathbf{T}_{n,k}^{1} = \left(ikP-u_{n-k,n} \right)^{-1}, & k' = k+1 \\

\mbox{(\ref{1b}), (\ref{1c}): } \mathbf{T}_{n,k}^{2} =  -\left(ikP-u_{n-k,n} \right)^{-1}, & k' = 0. \\
\end{array}
\end{equation}
We initialize the recursive procedure for the first column $n_0$ by
\begin{equation}
 \mathcal{C}^{n_0}_{0}(P,N) = \mathbf{T}_{0,0}^{i},
\end{equation}
with $i=1$ for a \emph{up} choice and $i=0$ for a \emph{low} choice.

The case of identical eigenstates is more complicated, and is explained in detail in \ref{same}.

We have thus constructed a method to compute the one body density matrix of our system, which we are going to apply in the next section to give some results.

\section{Results}\label{finalpart}

We now show, as an application, some results that are obtained through this method. The free parameters of our code are the number of particles $N$, the length of the system $L$, the interaction strength $c$, the Gaussian width $\Delta$, the number of strings we take in our superposed state $s$, and the spatial grid width $\delta x$. In the following, $n_0$ (giving rise to the centre-of-mass momentum $P_0=\pi n_0/L$) is taken equal to zero. The main quantities we are interested in from the density matrix are its eigenvalues $c_i$, which are independent of $n_0$.

The eigenvalues $c_i$ are defined through the  spectral decomposition
\begin{align}
\rho(x',x) = \sum_{i=0}^\infty c_i \phi_i^*(x')\phi_i(x) .
\end{align}
As the single-particle density matrix of \eqref{densityf} is positive semidefinite,  $c_i \ge 0$ and the eigenvalues sum up to the particle number $\sum_i c_i = \tr \rho = N$. If the largest eigenvalue (denoted by $c_0$) is significantly larger than all others, in particular when the number is macroscopically large, one speaks conventionally of the presence of a Bose-Einstein condensate \cite{Castin2001}. We will loosely call the ratio $c_0/N$ the condensate fraction, even if the condition for a Bose-Einstein condensate is not satisfied. A condensate fraction close to unity will indicate a quantum state that is well approximated by Hartree-Fock (or Gross-Pitaevskii) mean-field theory. While the ground state of $N$ attractive bosons is a single string state with vanishing condensate fraction for $L\to\infty$, it has been argued that the fully condensed Hartree-Fock ground state provides an insightful approximation \cite{Castin2001}. 

\begin{figure}[!h]
\centerline{
\subfloat[$N=6$, $c = -0.5$, $s=7$ and $\Delta = 0.05$]{
	\includegraphics[scale = 0.5]{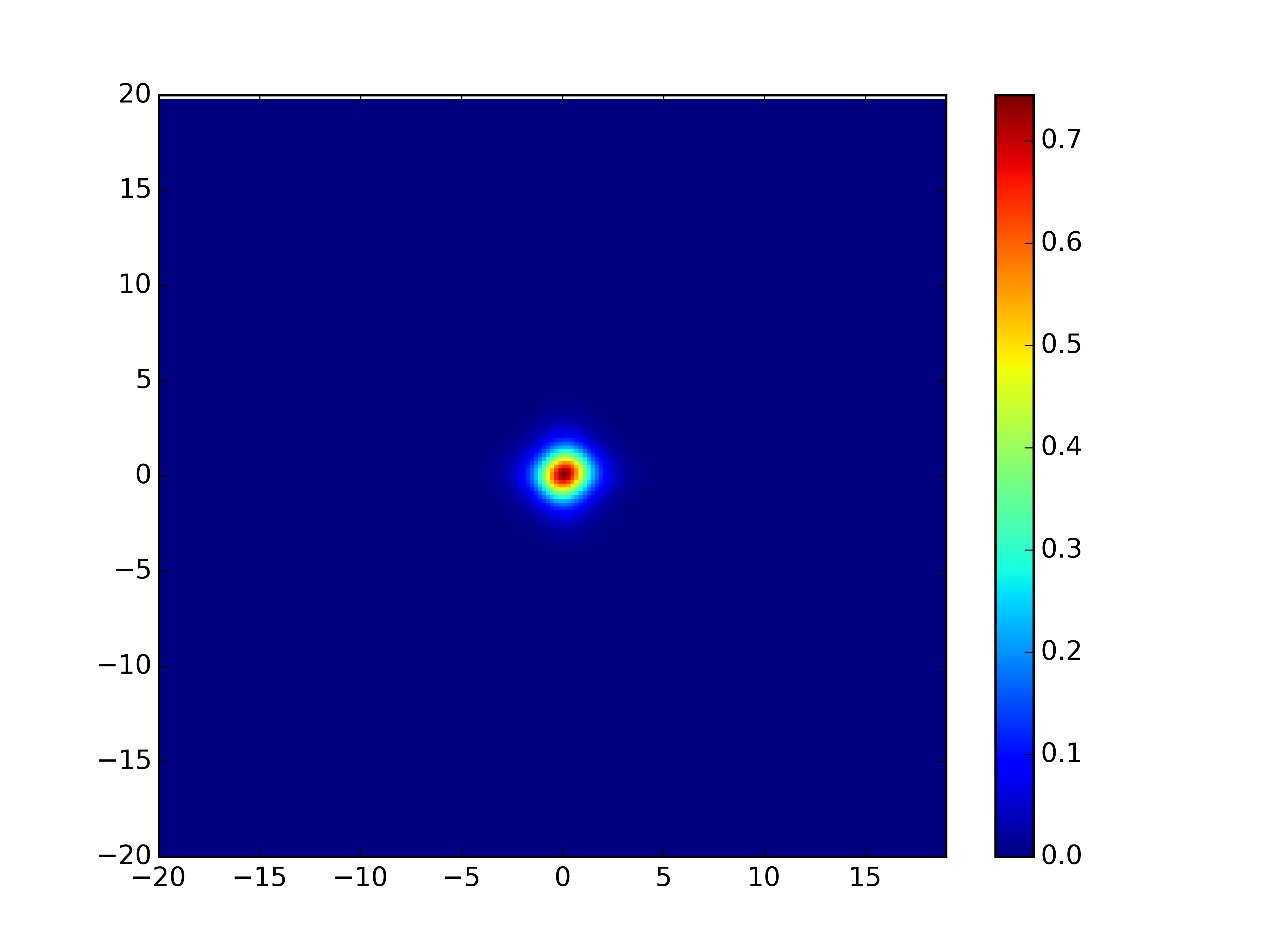}
	\label{premier}}
\subfloat[$N=10$, $c = -0.5$, $s=70$ and $\Delta = 20$]{
	\includegraphics[scale = 0.5]{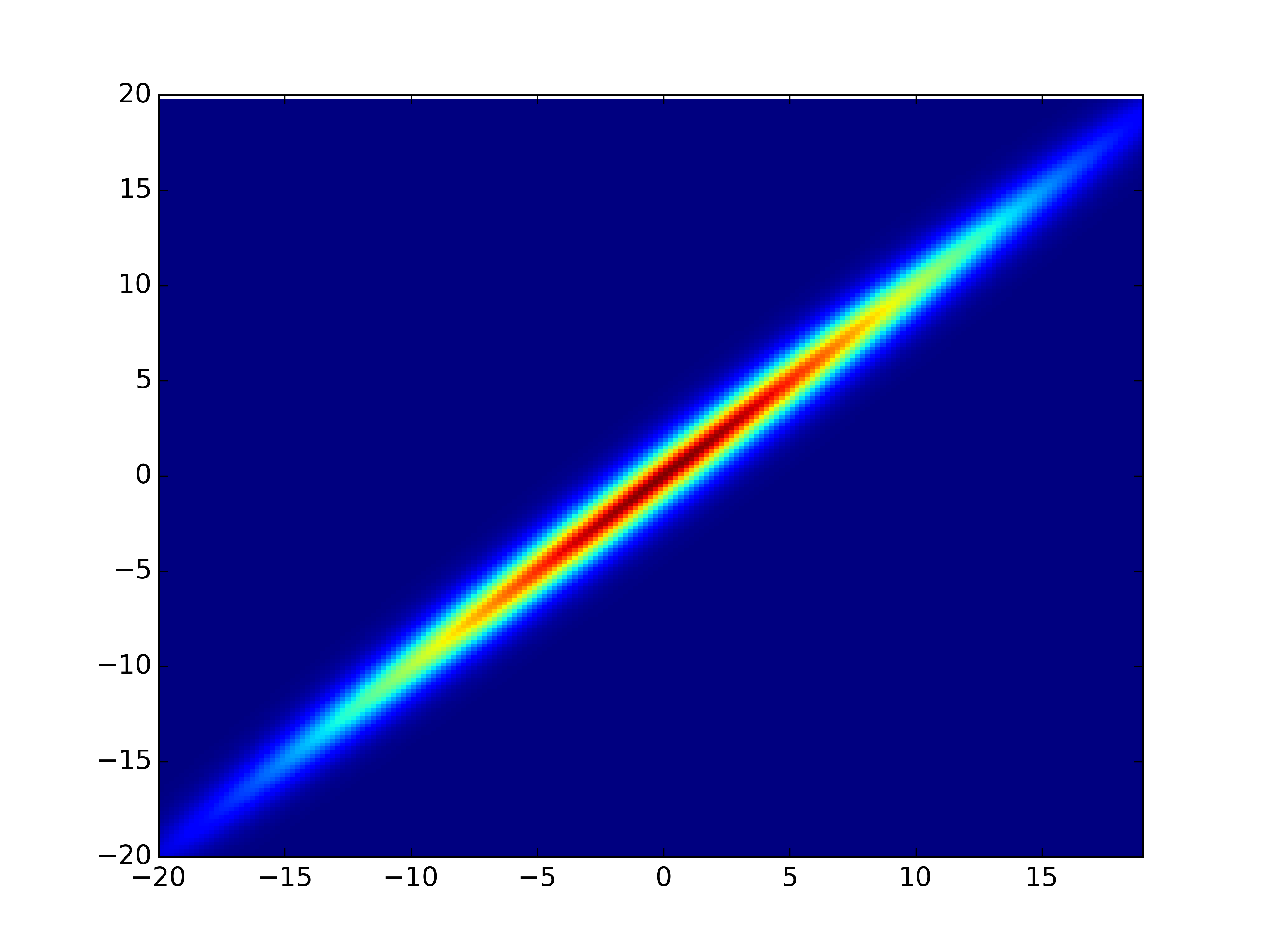}
	\label{deuxieme}}
}
\caption{Carpet plot of the single-particle density matrix \eqref{d-matrix} of a quantum soliton state. We can see that \ref{deuxieme} is more localized than \ref{premier}, resulting in a larger condensate fraction.} 
\label{onebodies}
\end{figure}

We can first look at the qualitative shape of the density matrix. As we can see from \fref{onebodies}, the density matrix is localized close to the diagonal. For a completely condensed state (i.e. when all the eigenvalues of $\rho$ are 0 except one), we expect the density matrix to 
have the four-fold symmetry of a product function, 
centred on zero. Indeed, for the product state describing a mean-field soliton, the matrix is of the form $\rho = \ket{G}\bra{G}$, with $\ket{G}$ being a state with a bell-shaped profile in position representation. The rounder and more elongated the matrix looks like, the more fragmented the condensate is, which means that other eigenstates beyond the dominant one become relevant. When we increase the number of particles keeping the other parameters fixed, we obtain a state that is more fragmented. Reversely, increasing the number of strings that we take in our superposition increases the condensation. Since $s$ should be as close to infinity as possible, it should be taken large enough to observe a saturation of the condensed fraction.

Restricting the threshold $s$ to a finite number may also lead to numerical artefacts. When we increase $\Delta$, we can reach a regime where the density is not localized anymore, and we see other peaks in the density appearing on the diagonal of the single-particle density matrix due to an insufficient sampling of the momentum-space distribution. This problem can again be avoided by increasing $s$.

The computational time needed to obtain a density matrix is a limiting factor:  on the NZIAS computer cluster (single CPU, Python implementation), for s = 10  and N = 6, it is of 2 hours, and it scales as $\mathcal{O}(s^2)$. The aim of this section is to present some preliminary results that highlight the relevance of the method, using modest computational resources.

\begin{figure}[!h]
\centerline{
\subfloat[]{
\includegraphics[scale = 0.5]{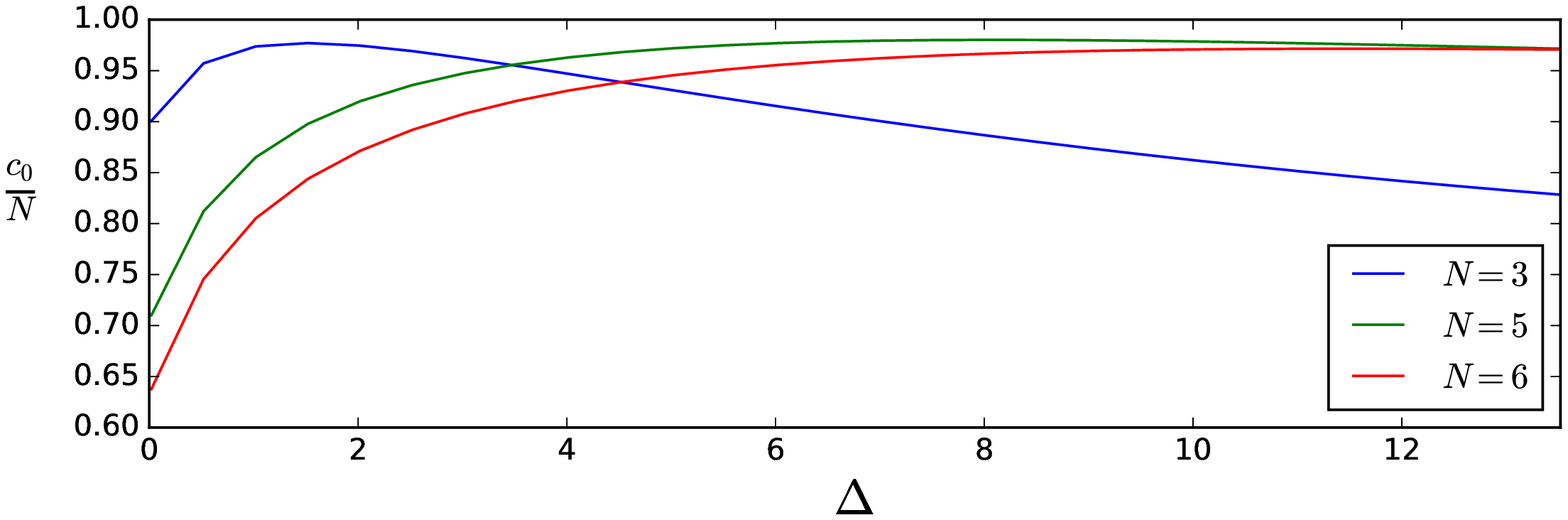}
\label{CvD}
}
}
\centerline{
\subfloat[]{
\includegraphics[scale = 0.5]{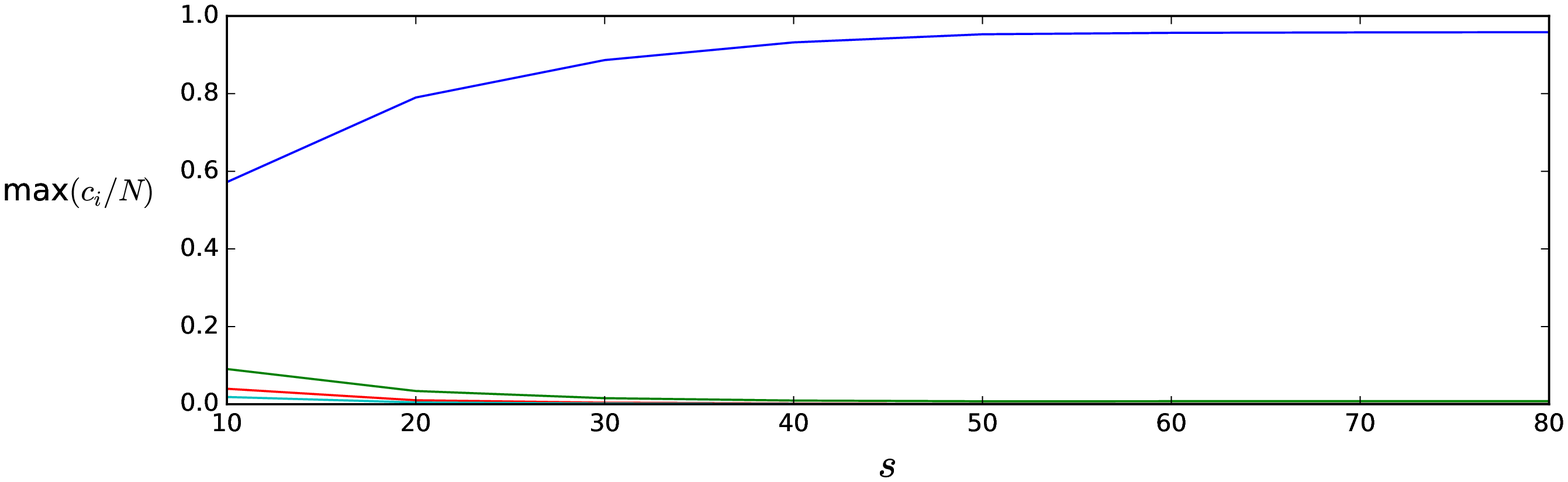}
\label{recursion}
}
}
\centerline{
\subfloat[]{
\includegraphics[scale = 0.5]{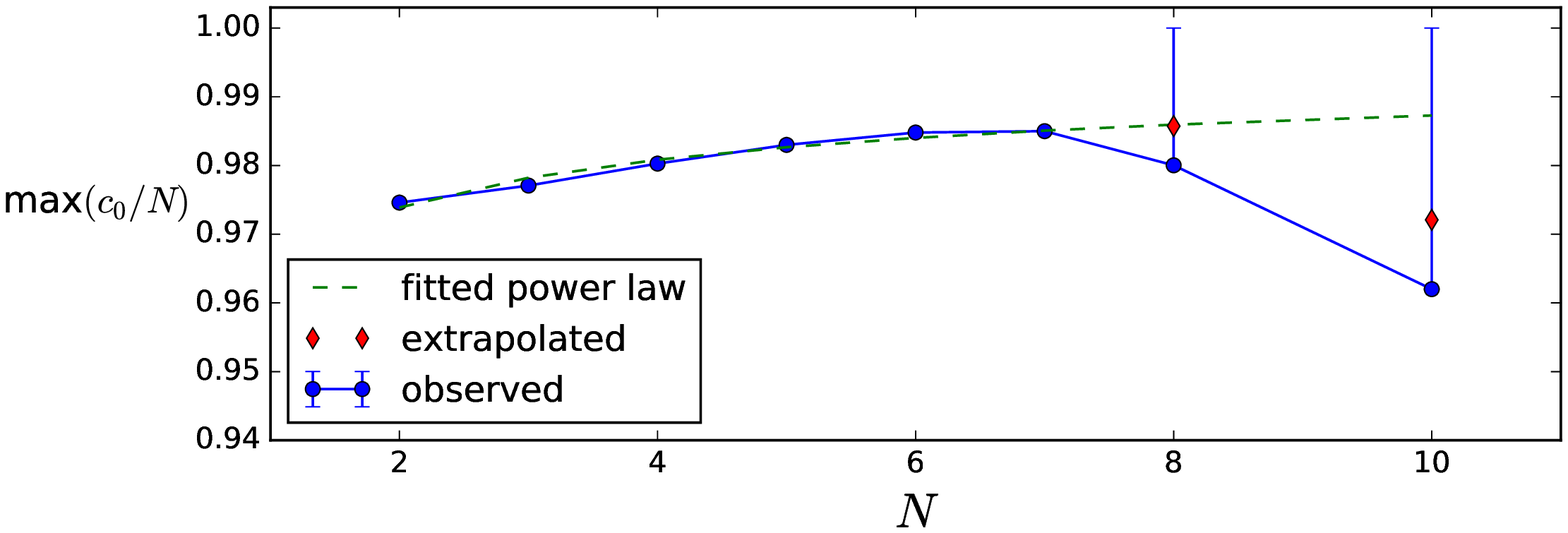}
\label{maxC}
}
}
\caption{Properties of the eigenvalues of the density matrix for $L = 25$, $c = -0.5$ and $\delta x = 0.3$. (a) Maximal eigenvalue of the density matrix versus the width of the state in momentum for different number of particles. (b) Four largest eigenvalues of the single-particle density matrix of the quantum soliton vs.\ the string number $s$ for $N = 6$. (c)   Maximal eigenvalue of the density matrix versus the width of the state in momentum for different number of particles (dots, and their associated error bars due to non optimized number of strings), extrapolated values from a least squares fitting (diamonds), and a fitted power law on points with no uncertainty (dashed line). We observe (a) that a maximum in condensed fraction is reached when $\Delta$ varies, (b) that this maximum increases with the number of strings and (c) that in the saturated regime of (b) the maximal condensed fraction scales as a power law with respect to $N$.}
\label{eigen}
\end{figure}

An interesting problem is to study the scaling of the maximal condensed fraction (with respect to $\Delta$, the spread in momentum of the state) with the number of particles.  Indeed, we see in \fref{CvD} that the condensed fraction has global maximum when $\Delta$ varies, that is reached for higher values of $\Delta$ as the number of particles increases. 

As mentioned above, it is necessary to optimize the number of strings to avoid sampling errors. Indeed, we see in \fref{recursion} that for a given particle number, the maximal condensed fraction becomes larger when the number of strings increases. Furthermore, the other eigenvalues of the density matrix go to zero as a consequence of the fact that we are approaching a Bose-Einstein condensate. From this figure we can determine an optimal number of strings to be used when studying the scaling problem, which saturates the condensed fraction. It is important to mention that this optimal value increases with the number of particles.

In \fref{maxC} we study the scaling of the maximal condensate fraction for up to ten particles.  We compute the density matrix for values of $\Delta$ from zero to 25. For up to seven particles, we are able to reach the maximum of condensed fraction within the considered range of $\Delta$.  For higher number of particles, we thus observe only a lower bound of the maximal condensed fraction, and we extrapolate\footnote{We fit the maximal eigenvalue data to $A + \sum_i B_i[1-\exp(-\Delta/\tau_i)]$ for $i \in \{1,2,3\}$, (using the least square criterion) in order to extrapolate a saturation value $A + \sum_i B_i$.} a saturation value, which is fraught with some uncertainty. 
Additional uncertainty for more than seven particles comes from the fact that number of strings (ranging from 70 to 120 strings) may not be large enough to  completely saturate the condensate fraction. 

The maximal condensed fraction is seen to very slowly grow with the particle number $N$. Given that repulsively interacting bosons in the Tonks-Girardeau gas regime have a condensate fraction that scales as $N^{-1/2}$ \cite{Lenard1966} we may anticipate some kind of power-law behaviour. We  have thus fitted  the power law
\begin{equation}
\mathcal{C}(N) = 1 - a N^{-\beta}
\end{equation}
for $N$ from one to 7 and obtained the coefficients $a = 0.04$ and $\beta =  0.44$. Although we cannot make a strong case for power-law scaling, from the available data at least a power law is feasible. A fully Bose-condensed soliton would thus be reached in the limit of large particle number. This is remarkably different from the Tonks-Girardeau gas where the condensate fraction tends to zero in the limit of an infinite system. Note that the maximal condensate fraction is independent of the interaction strength $c$, which provides a relevant length scale. Since, for large box size $L\gg 1/c$ it is the only remaining length scale, it can be scaled out of the problem and becomes irrelevant for the maximum of the condensate fraction.  

\begin{figure}[!h]
\centerline{
	\includegraphics[scale = 0.53]{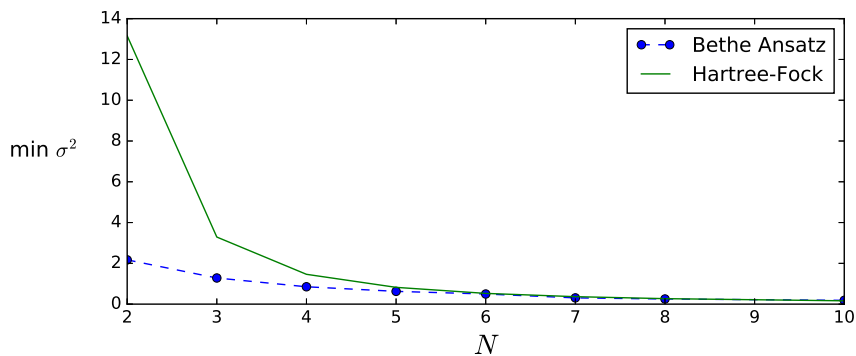}
	\label{maxvari}}
\caption{Spatial variance $\sigma^2 = N^{-1}\{\int x^2 \rho (x,x)dx  - [ \int x \rho(x,x) dx ]^2\}$  of the single-particle density versus particle number. Blue dots show the variance of the quantum soliton state $\ket{S}$ obtained as minimum over varying $\Delta$ for fixed   $s=120$, $L = 25$, $c = -0.5$ and the full (green) line shows the variance $\sigma_\mathrm{HF}^2$ of the Hartree-Fock solution of \eqref{eq:HFvariance}. Convergence of the quantum soliton to the mean-field solution is observed for large particle number.}
\label{vari}
\end{figure}

The length scale of the quantum soliton is considered  in \fref{vari}, which shows the  spatial variance of the number density $\sigma^2$ as a function of  particle number $N$ and compares it to the width of the Gross-Pitaevskii/Hartree-Fock soliton. It is seen that the mean-field soliton overestimates the width of the fundamental quantum soliton for small particle number but the two agree well for larger particle numbers.

The results of both \fref{maxC} and  \fref{vari} indicates that the quantum bright soliton asymptotically approaches the classical soliton for large particle numbers. A defining feature of the classical soliton is that it does not spread in time. In contrast, the quantum soliton $\ket{S}$,  has a centre-of-mass Gaussian wave function that obeys free-particle time evolution and spreads ballistically  \cite{Lai,Cosme2015a}. In the limit of large particle number $N$, the centre-of-mass corresponds to a heavy particle (with mass $\propto N$) and its ballistic expansion is slow compared to other relevant time scales \cite{Lai}.

\section*{Conclusion}\label{conclusion}

We have presented a method to compute the full single-particle density matrix of a quantum bright soliton, based on a numerical evaluation of a diagrammatic formalism.

As an application, we have computed a few relevant quantities, in particular the largest obtainable condensate fraction of a quantum bright soliton  for different number of particles, which has been found to be close to 0.97 for the range studied, and to increase with the number of particles. Furthermore, the spatial variance of the mean field soliton is recovered when the number of particles increases.

The interesting features revealed by results are a motivation for future studies of the time-evolution of the single-particle density matrix, which is straight forward with the presented approach. This will allow us to gain a deeper understanding of the fragmentation dynamics of quantum bright solitons and study the properties of higher-order solitons and soliton collisions. Furthermore, the diagrammatic formalism developed in this work is easily generalised to higher-order correlation functions such as two-particle densities. 

 \appendix

\section{Norm of the Lieb-Liniger eigenstates}\label{appen-norm}

In this appendix, we prove equation \eqref{Norm}. This norm is the one used for our numerical calculations. We want the eigenstates of the system to be normalized to one, and the trace of the single-particle density matrix to be $N$. When we compute the density function \eqref{densityf}, we discretise space on a grid of spacing $\delta x$. This means that the actual condition for the normalization of the density function is
\begin{equation}\label{A1}
\delta x\sum_{k = -L/\delta x}^{L/ \delta x} \rho(k \, \delta x,k \, \delta x)  = N.
\end{equation}
We can infer from this that, when computing the norm of the eigenstates, we are going to discretise only the last variable $x_N$, which is the only one that is not integrated in the definition of the density function, and on which we sum over in \eqref{A1}.

The condition for the normalization of the eigenstates is thus
\begin{multline}
\psh{p}{p} = \mathcal{N}^2 \sum_{k = -L/\delta x}^{L/\delta x} \delta x \int_{[-L,L]^{N-1}} dx_1 ... dx_{N-1} \\ \psh{p}{x_1,...,x_{N-1},k \, \delta x}\psh{x_1,...,x_{N-1},k \, \delta x}{p} = 1.
\end{multline}

As in \sref{part2}, we can first extend the integration domain to $[-\infty,\infty]^{N-1}$, and also, given the symmetry of the integrand, rewrite it in the fundamental domain $ -\infty \leq x_1\leq...\leq x_{N-1}\leq i \,\delta x \leq \infty$. We thus obtain
\begin{align}
\mathcal{N}^{-2} =& N!\, \delta x \sum_{k = -L/\delta x}^{L/\delta x} \int_{-\infty}^{k \, \delta x} dx_{N-1}... \int_{-\infty}^{x_2} dx_1 \nonumber \\ 
	&\;\;\;\;\;\;\;\times e^{c \sum (N-2j+1)x_j + c\,k\,\delta x} \nonumber \\
	=&N! \,\delta x \sum_{k = -L/\delta x}^{L/\delta x} 1 \prod_{j=1}^{N-1}\frac{1}{c(N-j)j} \nonumber \\
	=& \frac{N! \, \delta x}{c^{N-1}(N-1)!^2} \frac{2L}{\delta x}.
\end{align} 
Thus we recover \eqref{Norm}.

\section{Form factor involving identical eigenstates}\label{same}
 
In \sref{diff}, we explained how we can construct a diagrammatic representation that simplifies the computation of the integral $\mathcal{I}$ in the case of $P \neq 0$. Here we consider the case of $P = 0$. This case occurs when the argument of the exponential we are integrating is real, and is of the form $-u_{i,j}$ (see \eqref{udef} for the definition), and can vanish. It is the case when $j+i = N$. A very important fact to notice is that \emph{it can only vanish once per diagram}. If the argument of the exponential is different from zero, we can use the same representation as in \sref{diff}.

Let's assume that the cancellation happens for the integral on the variable $x_{n-1}$. Then the result of this integration will be $(x_{n} - x)$. When we compute the integral over $x_{m+1}$, we obtain
\begin{equation}
\int_x^{x_{n+1}} dx_{n} x_{n}e^{-u_{n,n}x_n} - x\int_x^{x_{n+1}} dx_{n} e^{-u_{n,n}x_n}.
\end{equation}

\begin{figure}[!h]
\centerline{
\subfloat{
 	\includegraphics[scale=0.35]{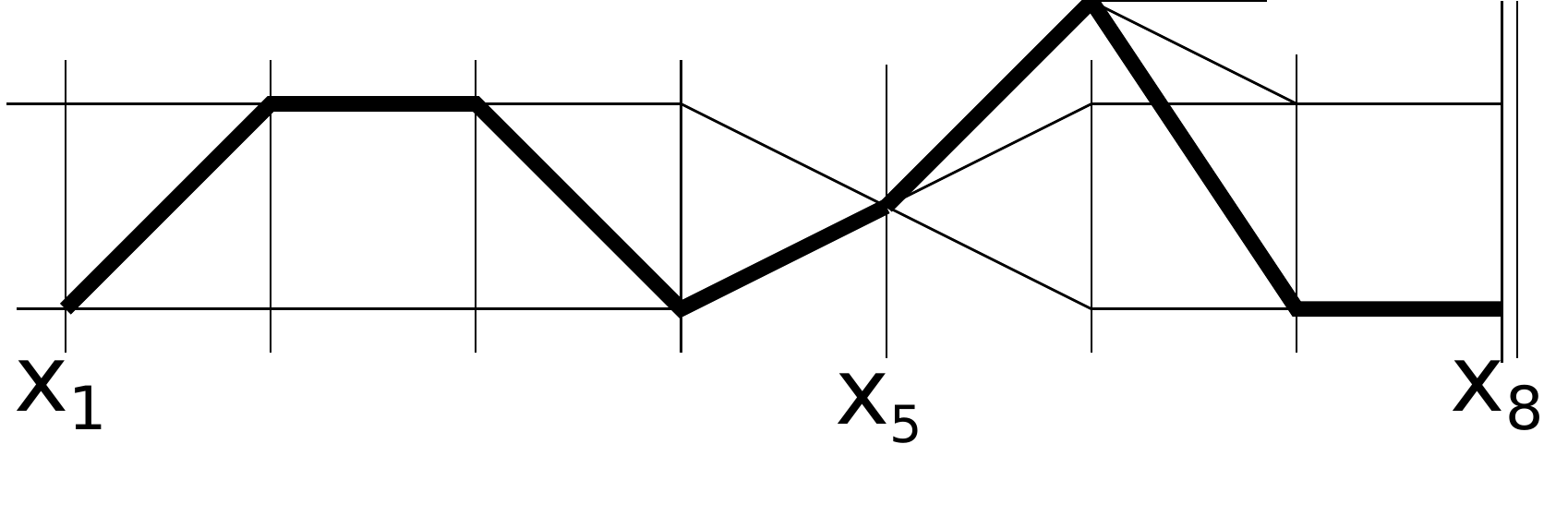}
}}

\centerline{
\subfloat{ 
	 \includegraphics[scale = 0.35]{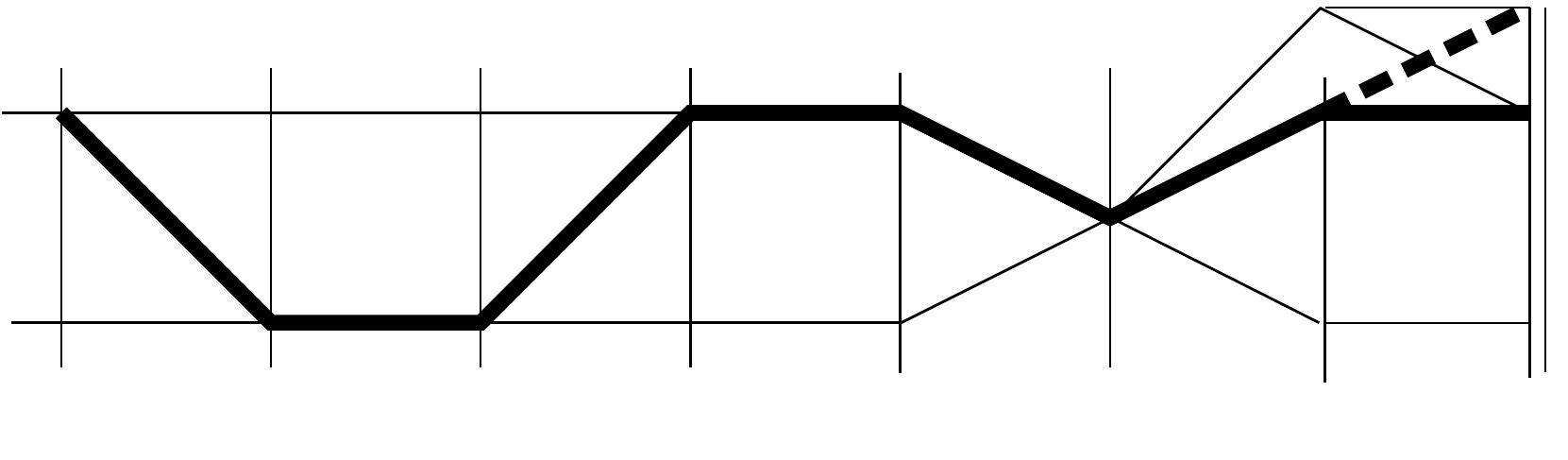}
}
}
\caption{Examples of bifurcations for the case N = 10. In the upper case, the cancellation of the exponential argument occurs when integrating $x_5$ (it is $u_{5,5}$) and in the second case for $x_6$ ($u_{4,6}$). These are the bifurcation points, where we have then three choices. The upper row cannot be reached by one of the lower rows, as it is the row where integrations by part are performed ($\alpha = 1$). The dashed line represents an example of a diagram that would not be allowed. }
\label{bifurcation}
\end{figure}

The second term follows the usual scheme of \ref{diff}, but the first term requires an integration by parts. After performing it, we get
\begin{equation}\label{exterms}
\frac{x_{n+1}}{u_{n,n}}e^{-u_n x_{n+1}} - \frac{1}{u_{n,n}^2}e^{-u_{n,n} x} + \frac{1-u_{n,n} x}{u_{n,n}^2} e^{-u_{n,n} x}. 
\end{equation}
Equation \eqref{exterms} contains three different types of terms, the first one will induce a new integration by parts in the next integral, the second one will act as a common \emph{up} choice in the diagram, and the third one as a \emph{low} choice (this is inferred from the exponential dependence it is carrying). What happens in this case is that the diagram faces a bifurcation: after the cancellation of the argument of the exponential, a new line of choices is added, as shown in \fref{bifurcation}. A consequence of the fact that $u$ can only vanish once per diagram is that we will only have one bifurcation per diagram, so we will never reach the case where we would have a double integral by parts to perform (with an integrand of the type $x^2e^{ux}dx$).

In the case where we are computing a form factor involving identical eigenstates, we have to change the recursion relation of \sref{diff} adding a new memory variable $\alpha \in \{0,1\}$ that stores the fact that an integration by parts is occurring. When constructing or reading the diagrams, we have two different regimes. The case where $\alpha= 0$ is computed in the exact same way than previously, using relations \eqref{diffrecursion}. The regime $\alpha=1$ is different and needs new relations. We give now the formula allowing to compute the contribution associated to a given diagram
\begin{equation}
 \mathcal{C}^{n+1}_{k',\alpha'}(P,N) = \mathbf{Q}_{n+1 ,k,\alpha}^{i} \mathcal{C}^{n}_{k,\alpha}(P,N) \text{.}
\end{equation}

\begin{figure}[!h]
\centerline{
\subfloat[]{
	\includegraphics[scale = 0.2]{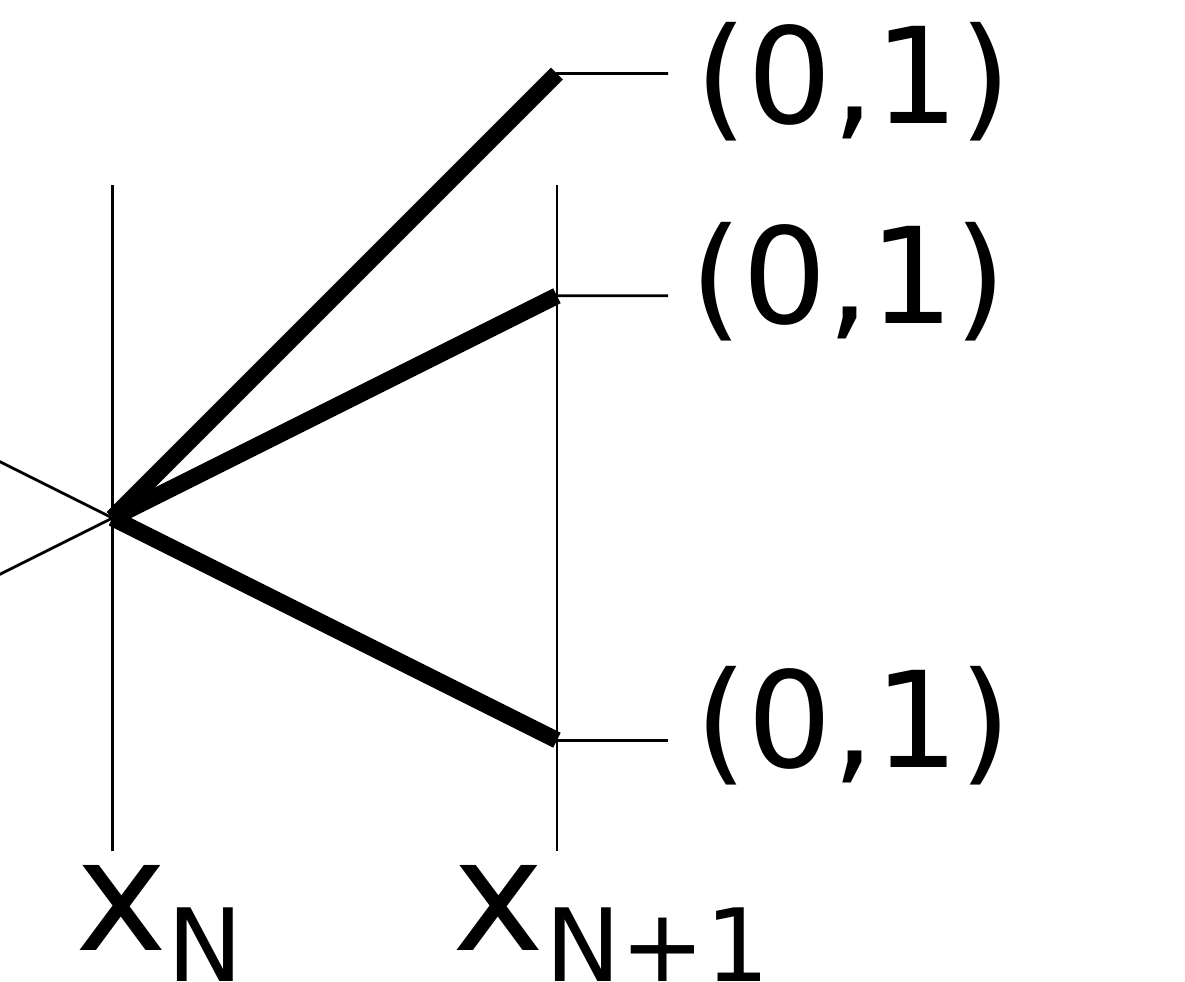}
	\label{2a}
} \;\;\;\;\;\;
\subfloat[]{
	\includegraphics[scale = 0.2]{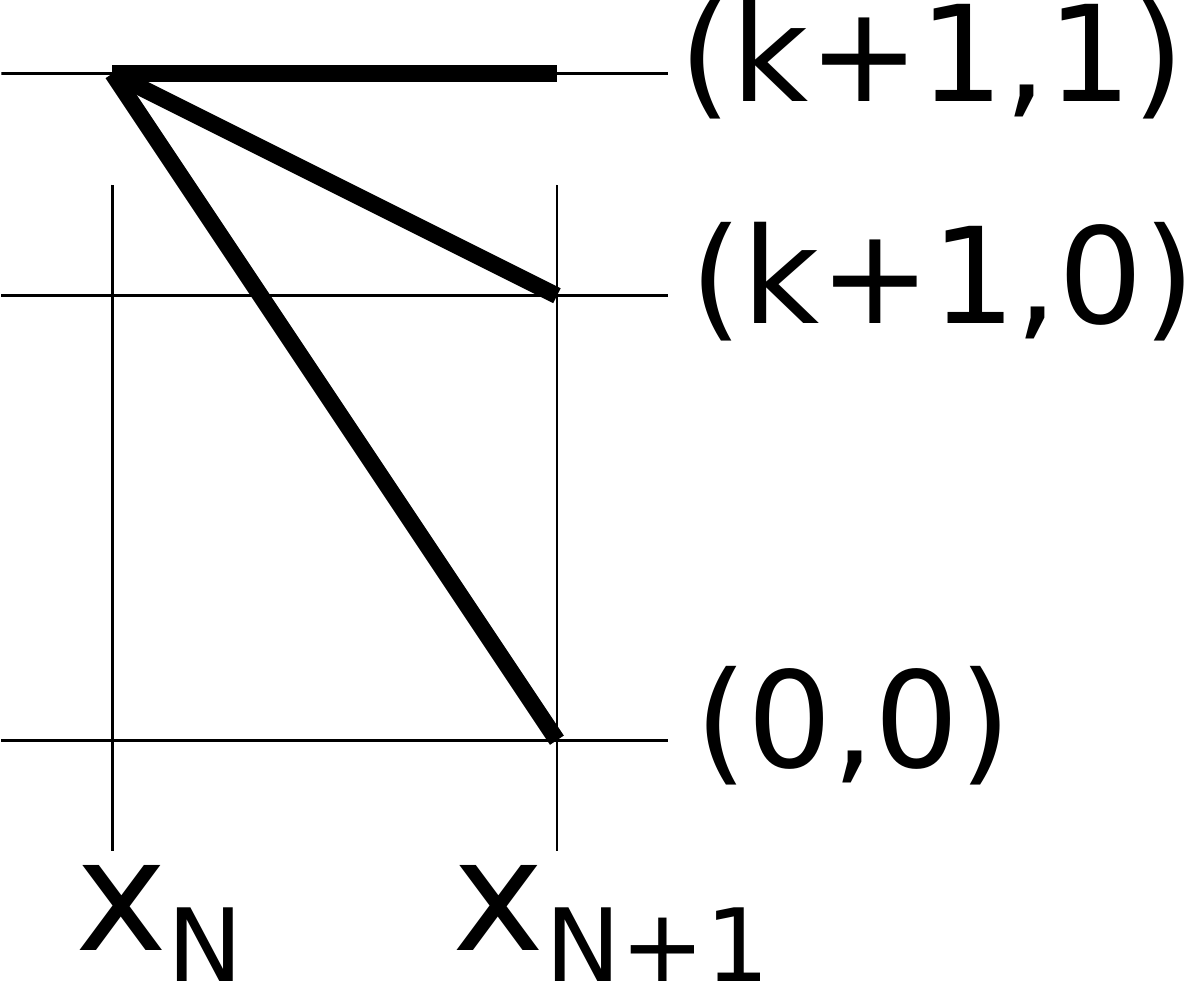}
	\label{2b}
}\;\;\;\;\;\;
\subfloat[]{
	\includegraphics[scale = 0.2]{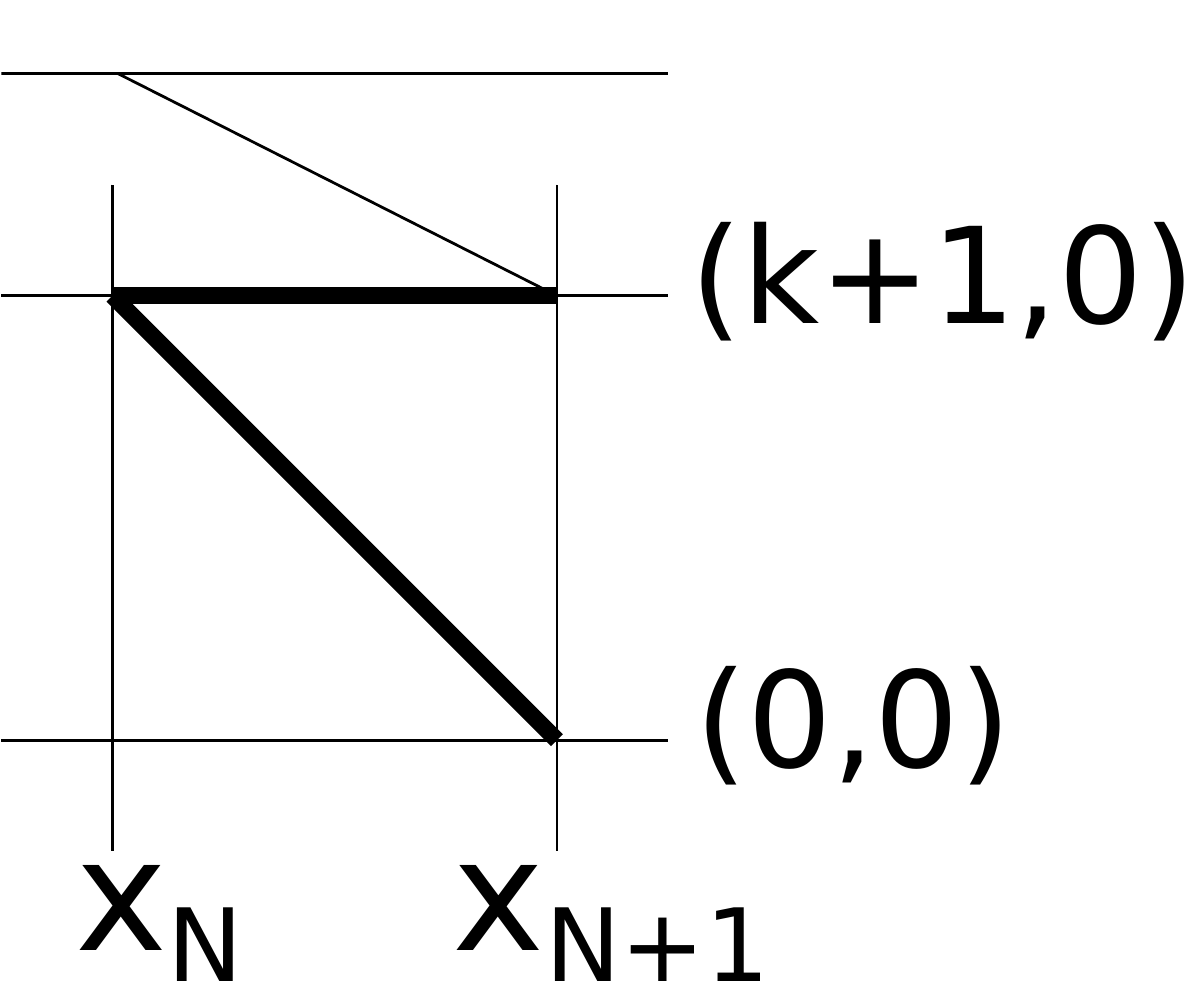}
	\label{2c}
}\;\;\;\;\;\;
\subfloat[]{
	\includegraphics[scale = 0.2]{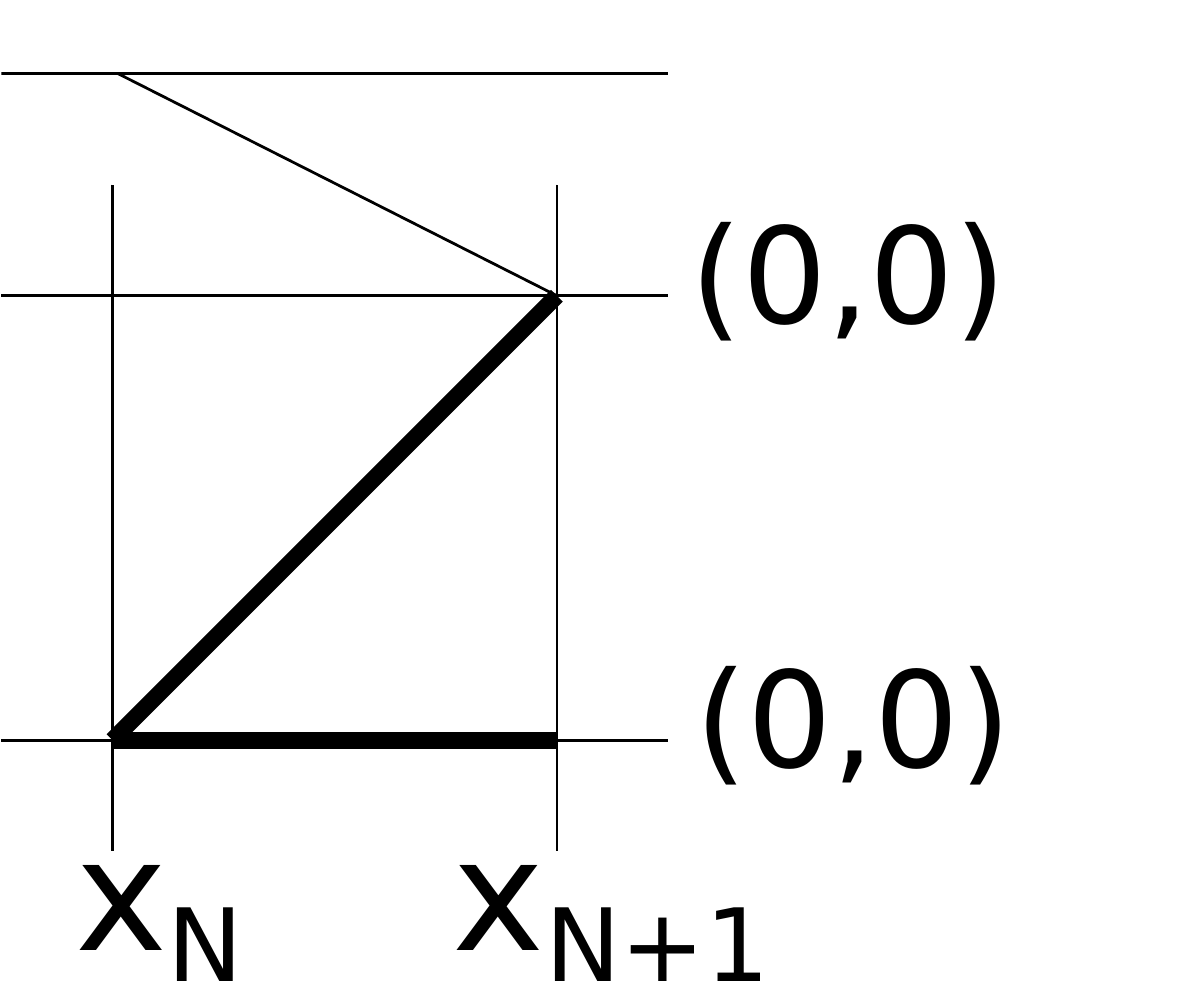}
	\label{2d}
}
}
\caption{Additional elementary diagrams for the case of a form factor between the same eigenstates. The diagram \ref{2a} is the bifurcation diagram, and the three other diagrams are the elementary diagrams in the regime $\alpha=1$. Bold lines represent the different possibilities in each case, associated with the couple $(k',\alpha')$ of new memory variables. When $\alpha=0$, refer to \fref{exdiag}.}
\label{exdiag2}
\end{figure}

The rules for the transition factor $\mathbf{Q}$ are associated  to the  elementary diagrams of \fref{exdiag} and \fref{exdiag2}. We give here these relations, and the couple of new memory variables $(k',\alpha')$:
\begin{equation}\label{eqrecursion}
\begin{array}{lll}
\mbox{(\ref{1a}), (\ref{1d}): }  \mathbf{Q}_{n,k,0}^{1} = \left(-u_{n-k,n} \right)^{-1},  \;(k+1,\alpha), \\
\mbox{(\ref{1b}), (\ref{1c}): } \mathbf{Q}_{n,k,0}^{2} = -\left(-u_{n-k,n} \right)^{-1}, \; (0,0), \\
\mbox{(\ref{2a}): } \mathbf{Q}_{n,k,0}^{3} = 1 , \; (0, 1), \; \mbox{only for}\; (2n-k) = N,  \\ 
\mbox{(\ref{2b}): } \mathbf{Q}_{n,k,1}^{1} = \left(-u_{n-k,n} \right)^{-1},  \; (\{0,k+1,\},\{0,1\}), \\
\mbox{(\ref{2c}): } \mathbf{Q}_{n,k,1}^{2} = \left(u_{n-k,n}^2 \right)^{-1}, \; (\{0,k+1\},0), \\
\mbox{(\ref{2d}): } \mathbf{Q}_{n,k,1}^{3} = \left(u_{n-k,n}x - 1\right)\left(u_{n-k,n}^2 \right)^{-1}, \;(0,0). \\
\end{array}
\end{equation}

In the above relations, it is important to add that $\mathbf{Q}_{n,k,0}^{3}$ can be used, and must be used only when the cancellation in the exponential occurs, which is the condition shown in \eqref{eqrecursion}, equivalent to $u_{n-k,n} = 0$.

As a final remark, we highlight that the complexity of this case is increased due to the fact that even the grid on which the diagrams are drawn can change according to the previous choices, between the regime $\alpha=1$ where it is a three-row grid and the $\alpha=0$ case where it has only two rows. The moment in which the bifurcation occurs depends on the previous choices of diagrams, but it can only occur once for a given diagram.


\section*{References}

\bibliographystyle{unsrt}
\bibliography{article}

\end{document}